\documentclass{raa}            

\usepackage{graphicx,times}             
\graphicspath{{fig/}}
\usepackage{natbib}
\usepackage{color}
\usepackage{amssymb,amsmath}
\usepackage{hyperref}
\hypersetup{pdftitle =paper, pdfauthor =G Li, pdfsubject=SEPs, pdfkeywords = SEP CME }
\hypersetup{colorlinks = true, linkcolor = blue, anchorcolor = red, citecolor = blue, filecolor = red, urlcolor = blue}

\newcommand{\beq}{\begin{equation}}
\newcommand{\eeq}{\end{equation}}
\newcommand{\beqa}{\begin{eqnarray}}
\newcommand{\eeqa}{\end{eqnarray}}

\begin{document}
\title{Probing satellite  galaxies in the Local Group by using FAST}
\volnopage{Vol. 0{200x} No.0, 000-000}
\setcounter{page}{1}
\author{Jing Li
     \inst{1,2}
 \and Yougang Wang \thanks{E-mail:wangyg@bao.ac.cn}
     \inst{2}
  \and Minzhi Kong \thanks{E-mail:kmz@hebtu.edu.cn}
      \inst{1}
   \and Jie Wang
       \inst{2}
    \and Xuelei Chen
        \inst{2}
     \and Rui Guo
          \inst{2}            }
 \institute{Department of Physics, Hebei Normal University, Shijiazhuang 050024, China \\
  \and                              
   Key Laboratory of Computational Astrophysics, National Astronomical Observatories, Chinese Academy of Sciences, Beijing, 100012 China}     

\date{Accepted XXX. Received YYY; in original form ZZZ}
\abstract{
The abundance of neutral hydrogen (HI) in satellite galaxies in the Local Group is important for studying the formation history of our Local Group. In this work, we generated mock HI satellite galaxies in the Local Group using the high mass resolution hydrodynamic  \textsc{apostle}  simulation. The simulated HI mass function agrees with the ALFALFA survey very well above $10^6M_{\odot}$, although there is a discrepancy below this scale because of the observed flux limit. After carefully checking various systematic elements in the observations, including fitting of line width, sky coverage, integration time, and frequency drift due to uncertainty in a galaxy's distance, we  predicted the abundance of HI in galaxies in a future survey that will be conducted by FAST.  FAST has a larger aperture and 
higher sensitivity than the Arecibo telescope. We found 
that the HI mass function could be estimated well around $10^5 M_{\odot}$ if 
the integration time is 40 minutes.  Our results indicate that there are 61 HI satellites 
in the Local Group, and 36 in the FAST field above $10^5 M_{\odot}$. This estimation is one order of magnitude better than the 
current data, and will put a  strong constraint on the formation history 
of the Local Group. Also more high resolution simulated samples are 
needed to achieve this target. 
\keywords{galaxies:Local Group, galaxies: mass function,methods:numerical}}


\label{firstpage}
\maketitle


\section{INTRODUCTION}
The cold dark matter (CDM) model is the preferred model in present cosmological studies. The observed properties of our Universe on cosmological scales are reproduced very well by the current standard $\Lambda\rm{CDM}$ cosmological model \citep{2012JPhCS.378a2012M}. However, on small scales there are three main problems. The first is  the ``missing satellite problem" \citep[e.g.][]{1993MNRAS.264..201K,1999ApJ...522...82K,1999ApJ...524L..19M},  in which the number of substructures in simulated dark matter halos, with the mass of the Milky Way (MW), is significantly larger than that of known luminous MW satellites. The second is the ``too-big-too-fail problem" \citep{2011MNRAS.415L..40B,2012MNRAS.422.1203B}, in which the count of subhalos with $V_{\rm max}> 25 {\rm km s^{-1}}$ predicted in MW-mass dark matter halos greatly exceeds that of observations. The third is the thin planar distribution of classical satellites. Most of the eleven brightest satellites of the MW are distributed in a plane that is roughly perpendicular to the MW disk \citep{1976RGOB..182..241K,1976MNRAS.174..695L,1982Obs...102..202L,1994ApJ...431L..17M,2000AJ....119.2248H},  while roughly half of M31's satellite galaxies belong to a vast, thin corotating plane
\citep{2013ApJ...766..120C, 2013Natur.493...62I}.

A large number of possible solutions to these problems have been proposed, and many of them rely on correcting the interplay between baryons and dark matter \citep[e.g.][]{2011MNRAS.417L..74D,2012MNRAS.422.1231G,2012ApJ...761...71Z,2013ApJ...765...22B,2014ApJ...786...87B}. Another possibility is that some groups have accreted less massive subhalos in their formation history and display a gap between their satellites  \citep{2016MNRAS.460.2152K}. The third perspective on these problems is that a large fraction of satellites have been missed in observations. In other words, it is possible that some dark matter dominated galaxies exist with no baryons and /or they have baryons, but no stars \citep{1997A&A...328L..25H,1997MNRAS.292L...5J}. These cases are called dark galaxies. Although some dark galaxies have no star formation, they still contain proportions of ionized atomic and molecular hydrogen, which could be potentially detected via a blind 21-cm survey of the sky, and  extensive studies have been performed in discussing the possible existence of dark galaxies \citep{1997MNRAS.292L...5J,2002MNRAS.336..541V,2005ApJ...634.1067T,2006NewA...12..201D,2007math......2034K,2007ApJ...670.1056M}.

Objects with neutral hydrogen devoid of stars have been known for many years. This kind of object contains high velocity clouds \citep{1997ARA&A..35..217W,1999AAS...19510202B,2000math......7069B,2001MNRAS.323..904L,
2005ApJ...623..196C,2007ASSP....3..319P}, the Leo Ring \citep{1983BAAS...15..876S, 2003ApJ...591..185S}, and clouds close to bright galaxies \citep{2001MNRAS.328..277B, 2001ApJ...555..232R, 2002AAS...201.5801L}. However, none of these objects have characteristics of a galaxy, for exmaple the HI emission region is smaller than the scale of a glaxy and the velocity is not greater than a few tens of km $s^{-1}$. Although several candidates have been claimed, such as VIGOHI 21\citep{2005AAS...20718813M, 2007ApJ...670.1056M}, HVC Complex H \citep{2005ApJ...621..757S} and Local Group galaxy LGS3 \citep{2002ApJ...580L.129R}, none of them has been proved. Therefore, using ongoing and future radio telescope projects to find these dark galaxies is a very important effort.

One import HI survey is the Arecibo Legacy Fast ALFA (ALFALFA)  survey, which is the second generation blind extragalactic HI survey exploiting Arecibo's superior sensitivity, angular resolution and digital technology to conduct a census of the local HI Universe over a cosmologically significant volume \citep{2005AJ....130.2598G}. Some exciting results have been discovered by the ALFALFA  survey \citep[e.g.][]{2007ApJ...665L..15K}. A future HI survey can be carried out by the Square Kilometre Array (SKA), which will have very high angular resolution and sensitivity. Before the full SKA comes into operation, another large radio telescope, the Five-hundred-meter Aperture Spherical Telescope (FAST), has great potential for HI surveys. FAST is the largest single aperture radio telescope ever built and the construction was finished in 2016 \citep{2011IJMPD..20..989N}. The sensitivity of FAST is at least 2.5 times that of the Arecibo telescope, i.e., more faint objects can be observed
by FAST.  Another advantage of FAST is that there are 19 beams in the L-band, which increases  the survey speed significantly. In this paper, we will forecast the potential of detecting faint galaxies in the Local Group via 21-cm radiation by using FAST.


\section{simulations}
The simulated data we use in this paper are from the \textsc{apostle}  (A Project Of Simulating The Local Environment)  simulation project \citep{2016MNRAS.457.1931S,2016MNRAS.457..844F}.
The \textsc{apostle} simulations are a suite of cosmological hydrodynamic simulations of twelve Local Group-like environments selected to match the kinematics of Local Group members.
Each selected volume has been re-simulated at several resolutions, both as dark matter only, and as hydrodynamic simulations,  with the code developed for the Evolution and Assembly of GaLaxies and their Environments (\textsc{eagle}; \citealt{2015MNRAS.450.1937C,2015MNRAS.446..521S}) project.

The \textsc{eagle} code is a modified version of \textsc{p-gadget-3}, which itself is an improved version of the \textsc{gadget-2} code \citep{2005MNRAS.364.1105S}. The \textsc{eagle} code includes star formation, feedback from evolving stars and supernovae, metal enrichment, cosmic reionization, and the formation and energy output from supermassive black holes/active galactic nuclei. The details of this code are described in \cite{2015MNRAS.446..521S} and \cite{2015MNRAS.450.1937C}.  The $z=0.1$ stellar mass function and galaxy sizes from $10^8$ to $10^{11}M_{\odot}$ in a cosmological volume of $100^3\rm{Mpc^3}$ can be successfully reconstructed. Many other properties and scaling laws of observed galaxy populations, such as evolution of the stellar mass function
\citep{2015MNRAS.450.4486F} and the luminosities and colours of galaxies \citep{2015MNRAS.452.2879T}, also can be reproduced in the   \textsc{eagle} simulation.

The  \textsc{apostle} simulations adopt the same parameters as used in the $100^3\ \rm{Mpc^3}$ L100N1504 \textsc{eagle} reference simulation \citep{2015MNRAS.446..521S}. In the  \textsc{apostle} simulations, twelve Local Group environments are zoom simulations, and are selected from a dark matter only simulation, which includes  $1620^3$ particles in a box size of 100Mpc on each side using the \emph {Wilkinson Microwave Anisotropy Probe \rm{(}WMAP-7\rm{)}} cosmology \citep{2011ApJS..192...18K}.  In order to match the kinematics of Local Group members, each selected volume contains a pair of halos in the virial mass range $5\times10^{11}-2.5\times 10^{12}M_{\odot}$, with median values of $1.4\times 10^{12} M_{\odot}$ for the more massive halo and $0.9\times 10^{12} M_{\odot}$ for the less massive halo. In addition, the distance between two halos is $800\pm 200$ kpc, and the two halos have radial velocity 0-250 $\rm {km\ s^{-1}}$  and tangential velocity smaller than $100\ {\rm km\  s^{-1}}$.  It also requires that there is no additional halo larger than the smaller of the pair within 2.5Mpc from the midpoint of the pair. More details about the selection criteria of the Local Group can be found in \cite{2016MNRAS.457..844F}.

The resolutions of  the \textsc{apostle} resimulations are labeled L1, L2, and L3, which refer to high, medium, and low resolution, respectively. Among the 12 selected pairs, only the AP-1 and AP-4 simulations have the L1 resolution, and the gas resolution in AP-4 is two times  that in AP-1. Moreover, the most massive halo in AP-4 is in a better virial equilibrium state than that in AP-1. Therefore, we use the AP-4 simulation with the L1 resolution.
In the AP-4 simulation with L1 resolution, the mass of each gas particle is $\mathrm{0.49\times10^4}$$M_\odot$.  The separation between the pairs is 790 kpc. The more massive halo is ${1.38\times10^{12}M_\odot}$ and the less massive halo is  ${1.35\times10^{12}M_\odot}$.  The relative radial and tangential velocities  of the pair are  $\rm{-59\ km\ s^{-1}}$ and $\rm {24\ km\ s^{-1}}$, respectively  (See Table 2. in  \citealt{2016MNRAS.457..844F}).

\begin{figure}
\centering
\includegraphics[width=6cm]{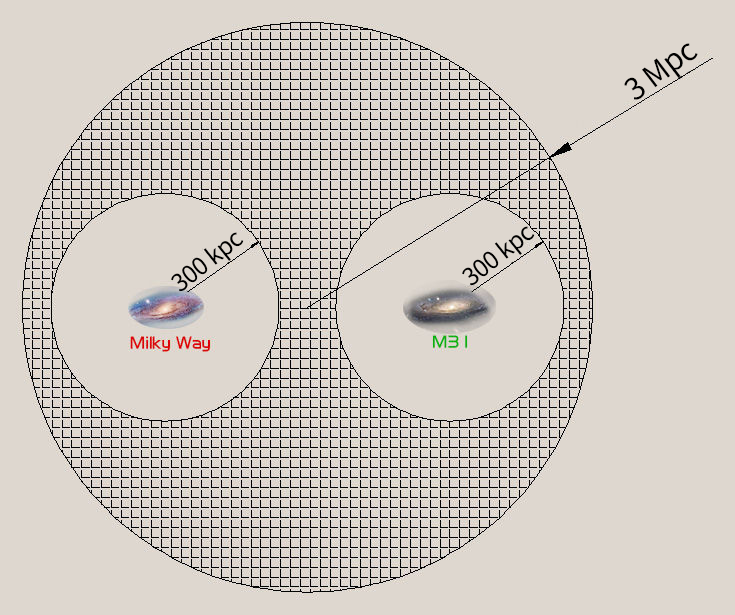}
\caption{A diagram of the mock survey region. A spherical region within 3Mpc from the midpoint between the MW and M31 is selected. Two small spherical regions centered on M31 and the MW with radius 300kpc are removed.}
\label{fig:local_region}
\end{figure}

Our observed objects are the satellites in the Local Group. The Local Group region is defined as follows. A  sphere with radius 3Mpc  centered on the mid point of the pairs has been selected for our studies.  Although 3Mpc is larger than the typical Local Group region, the resolution within 3Mpc around the middle point of the pairs in the  \textsc{apostle} simulation is high,  and we adopt a sphere with radius of 3Mpc as the Local Group to include more satellites. We take the more massive as corresponding to M31, and the less massive halo as corresponding to the MW. In order to avoid satellites in the MW and M31, two spheres centered on the MW and M31 halos with radius 300kpc are removed, which can been seen from Figure~\ref{fig:local_region}. There are a total of 61 satellites with HI mass larger than $10^5M_{\odot}$ in our selected region of the Local Group.

\section{THE FAST telescope}

FAST is a Chinese mega-science project that has  built the largest single dish radio telescope in the world. FAST also represents a Chinese contribution in the international effort to prepare for the Square Kilometer Array (SKA)  \citep{2011IJMPD..20..989N, 2013IAUS..291..325L}.

FAST is similar to the Arecibo telescope, but with a  larger aperture of 500 meters.  It is located in a deep karst depression in Dawodang, Guizhou province in southwestern China at  a latitude of about 26 degrees. This allows the facility to observe up to 40 degrees away from
the zenith angle without a notable loss in gain.  The construction of FAST was finished in September 2016.

Thanks to being the largest single dish radio telescope, FAST has high sensitivity, which enables the project to achieve some important goals, such as surveying neutral hydrogen in the MW and other galaxies, detecting faint pulsars, looking for the first shining stars, hearing possible signals from other civilizations, etc.

\par

FAST covers frequencies from 70 MHz to 3GHz, and the L-Band is 1.05-1.45 GHz. Since there are 19 beams in the L-Band, the HI survey speed will be improved significantly.  The main parameters describing FAST are listed in Table 1 of \cite{2011IJMPD..20..989N}.
The main reflector of FAST is a spherical cap with radius of $300\ {\rm m}$ and an open up to $500\ {\rm m}$ in diameter.
The aperture efficiency of 70$\%$ gives $\rm{A_{eff}=50\ 000\ m^2}$. At 1.4 GHz (the 21cm HI line), the system temperature is $\rm{T_{sys}=25K}$. The thermal noise for an observing time $t$ and a frequency bandwidth of $\triangle\nu$ can be computed by \citep{2008MNRAS.383..150D}
\begin{equation}
\sigma_{\rm noise}=\sqrt{2}\frac{kT_{sys}}{A_{eff}}\frac{1}{\sqrt{\Delta\nu t}}
\end{equation}
where $k=\rm{1380\ Jy\ m^2\ K^{-1}}$ is the Boltzmann constant.
\par
The flux limit for a specific signal-no-noise ratio(S/N) is given by \citep{2008MNRAS.383..150D}
\begin{equation}
S_{\rm lim}=(S/N)\frac{\sigma_{\rm noise}}{1+z}.
\end{equation}
A given galaxy will be detected if its flux is lager than $S_{\rm lim}$.

\section{The HIMF and stellar mass function}
\begin{figure}
\centering
\includegraphics[width=11cm]{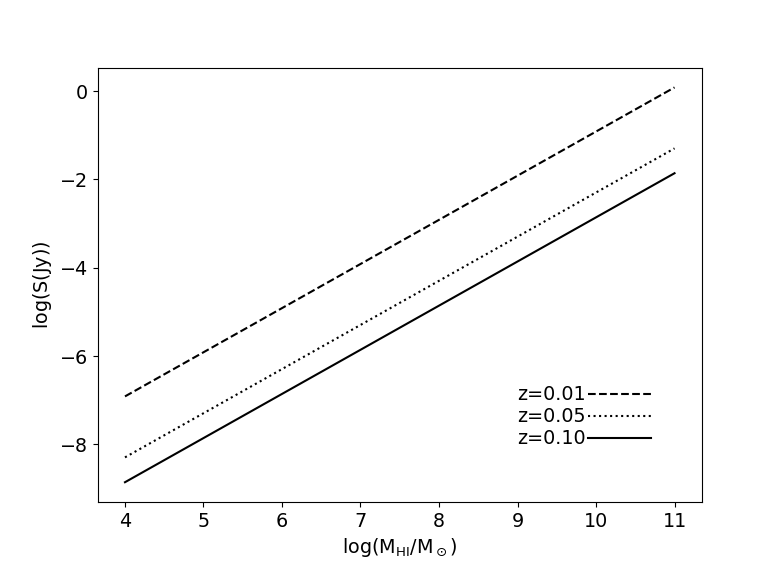}
\caption{The dependence of the HI flux on the galaxy HI mass with redshift z=0.01 $(dashed)$, 0.05 $(dotted)$  and 0.1 $(solid)$ lines. The linewidth $\Delta V_0$ is assumed to be 200 $\rm{km~s^{-1}}$.}
\label{fig:flux}
\end{figure}

For a galaxy at redshift z, the HI mass is given by \citep{1975gaun.book..309R}

\begin{equation}
\frac{M_{\rm HI}}{\rm M_{\odot}}=\frac{2.35\times10^5}{1+z}\bigg[\frac{d_{\rm
 L}(z)}{\rm Mpc}\bigg]^2\bigg(\frac{S}{\rm
Jy}\bigg)\bigg(\frac{\Delta V_0}{\rm km\ sec^{-1}}\bigg)
\label{eq:MHI}
\end{equation}
where $d_{\rm L}(z)$ is the luminosity distance to the galaxy, $S$
is the observed flux, and $\Delta V_0$ is the linewidth. Here we assume that the
bandwidth is $\Delta\nu=1 {\rm MHz}$, then the corresponding velocity
linewidth is ~$200\ \rm{km~sec^{-1}}$. Figure~\ref{fig:flux} shows the
dependence of the observed flux on the  HI mass for different
redshifts.

Research on the HI mass function (HIMF), defined as the number density of HI as a function of HI mass at the present epoch, was pioneered by \cite{1990AJ....100..999B} as  a diagnostic tool for
estimating completeness of optical galaxy catalogs. Measuring the HIMF is also important for understanding the formation and evolution of galaxies. The HIMF is usually described by the Schechter function \citep{1976ApJ...203..297S}
\begin{equation}
\Theta(M_{\rm HI}){\rm d}M_{\rm HI}=\theta^{\ast}\bigg(\frac{M_{\rm HI}}{M_{\rm HI}^{\ast}}\bigg)^{\alpha}\exp\bigg(-\frac{M_{\rm HI}}{M_{\rm HI}^{\ast}}\bigg){\rm d}\bigg(\frac{M_{\rm HI}}{M_{\rm HI}^{\ast}}\bigg)
\label{eq:HIFM}
\end{equation}
where $\theta^{\ast}$, ${M_{\rm HI}^{\ast}}$ and $\alpha$ are the fitting parameters.

\begin{figure}
\centering
\includegraphics[width=11cm]{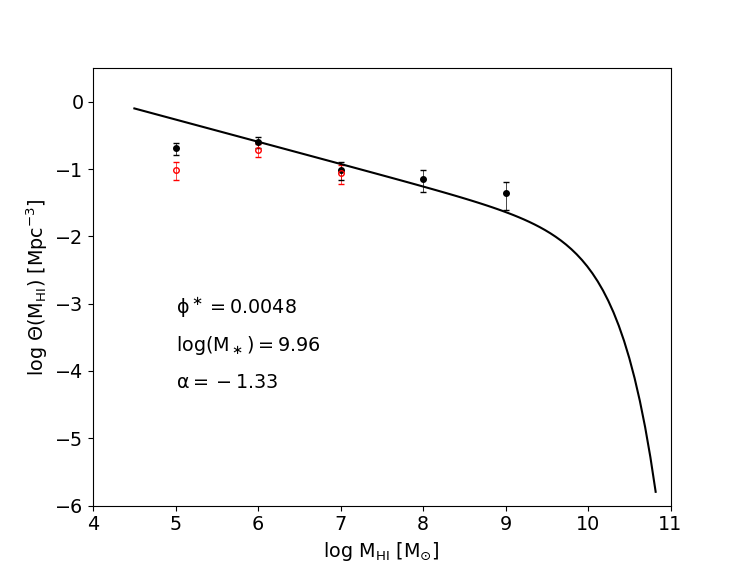}
\caption{The HIMF function of galaxies in the Local Group from our simulation data (black filled circles with error bars ) and from the ALFALFA Survey \citep{2010ApJ...723.1359M} (black solid line). The corresponding parameters in the Schechter function (black solid line) are labeled in the bottom left of the panel. The red circles with error bars show the HIMF function of galaxies within 300 kpc of both the MW and M31.}
\label{fig:HIMF}
\end{figure}

In Figure~\ref{fig:HIMF}, we show the HIMF for satellites in the Local Group from our simulation data (black filled circles with error bars) compared with that from the  ALFALFA Survey \citep{2010ApJ...723.1359M} (black solid line). The error bars on the dotted line are estimated by Poisson noise.  It is found that the HIMF in our simulation has a good consistency with that of the ALFALFA survey  if the HI mass  of the satellites is larger than $10^6M_{\odot}$.
At the low mass end, our result is lower than that predicted by the ALFALFA survey. For comparison, the HIMF of galaxies within 300 kpc of both MW and M31 is also given in Figure ~\ref{fig:HIMF} by red open circles, and the result is similar to the HIMF of Local Group galaxies. Limited by the sensitivity of the Arecibo telescope, the ALFALFA survey can only detect a galaxy with HI mass larger than $10^{6.2}$. Since FAST has higher sensitivity than the Arecibo telescope, galaxies with HI mass  lower than $10^{6.2}$ can be detected, which can help us to test the validity of the simulation result at the low mass end of the HIMF.

\begin{figure}
\centering
\includegraphics[width=11cm]{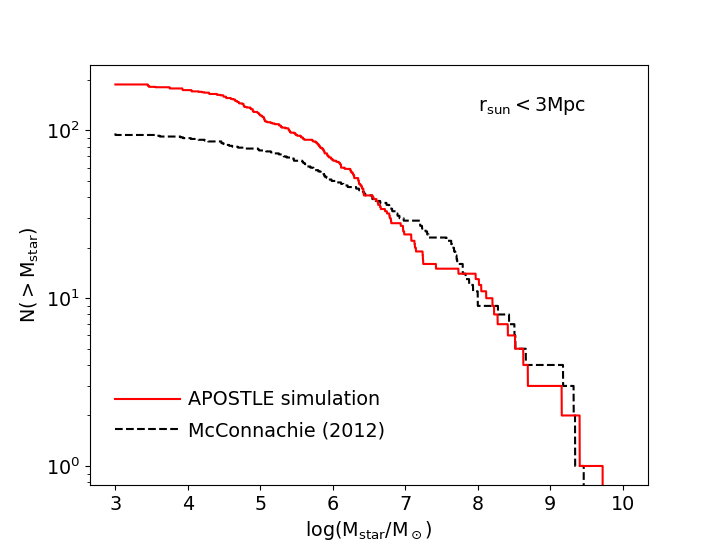}
\caption{Stellar mass functions in the \textsc{APOSTLE} AP-4 simulation within 3 Mpc of the Sun (red solid line) compared to the observed stellar mass function (black dashed line). The observational result is from \cite{2012AJ....144....4M}. }
\label{fig:smf}
\end{figure}

Figure~\ref{fig:smf} shows the galaxy stellar mass function in the \textsc{APOSTLE} AP-4 simulation  within 3 Mpc of the Sun (red solid line), which has the same volume as the observation in \cite{2012AJ....144....4M}. The observed stellar mass function is also plotted in this figure by the black dashed line.  It is seen that the stellar mass function in the \textsc{APOSTLE} AP-4 simulation is consistent with that of observations if the stellar mass is larger than $10^{6.5}M_{\odot}$. At the low stellar mass end, the stellar mass function in the simulation is larger than that of observations. This result was also found in \cite{2014MNRAS.438.2578G}. This difference may be due to incompleteness of the observations at the low stellar mass end \citep{2014MNRAS.438.2578G} or subgrid physics in the \textsc{APOSTLE} simulation \citep{2016MNRAS.457.1931S}. Among the 61 satellites in the Local Group, 12 satellites have not stars, which are  the ``dark" minihalos \citep{2017MNRAS.465.3913B}. The detailed properties and the observations of these minihalos have been discussed in \cite{2017MNRAS.465.3913B}. It is possible that FAST could identify some of these objects.

\begin{figure}
\centering
\includegraphics[width=11cm]{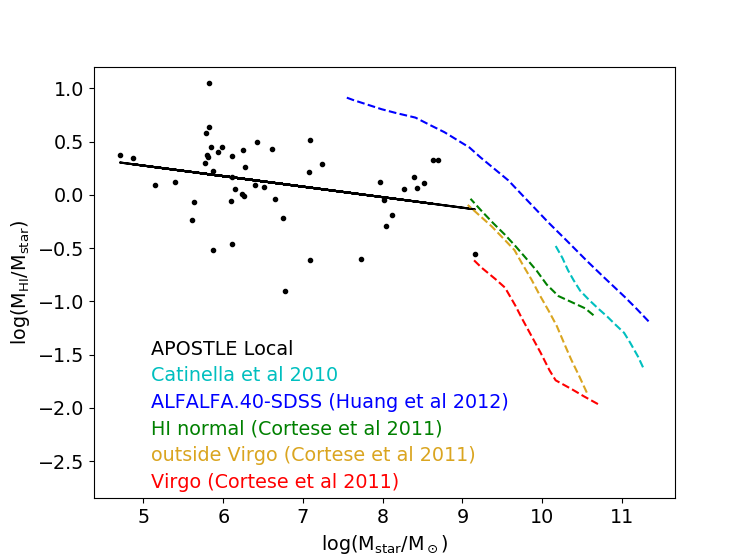}
\caption{Relation between HI mass and stellar mass. The black dots and the black solid line are the results from the \textsc{APOSTLE} AP-4 simulation, and the other dashed lines are the observation results. The cyan dashed line is the result from the GASS galaxies with $M_{\rm star}>10^{10}\ M_{\odot}$ \citep{2010MNRAS.403..683C}. The blue dashed line represents the result from the ALFALFA galaxies \citep{2012ApJ...756..113H}. The green, yellow and black dashed lines represent the results from \cite{2011MNRAS.415.1797C} for galaxies with normal HI, those outside the Virgo Cluster and those inside the Virgo Cluster, respectively.}
\label{fig:MHI_Mstar}
\end{figure}

In Figure~\ref{fig:MHI_Mstar}, we show the relationship of the HI mass with stellar mass. The black dots are the results from the  \textsc{APOSTLE} AP-4 simulation. The black solid line indicates the following equation   
\begin{equation}
 \log(M_{\rm HI}/M_{\rm star})=-0.0988\log(M_{\rm star})+0.769.
\end{equation}
The observed results are also shown in this figure. It is seen that the relation between the HI mass and the stellar mass in the simulation is different from those in the observations. Our result is in a good agreement with that from a normal HI galaxy sample, and an HI galaxy outside the Virgo Cluster individually is around $10^9M_{\odot} $, but we cannot check above this scale since we do not have simulations of more massive satellites.

\section{linewidth of  galaxies}
From equation~\ref{eq:MHI}, we know that the observed HI flux is correlated to the linewidth of galaxies. Therefore, it is important to estimate the linewidth of the galaxies.
Since our galaxies are from the simulation, we can determinate the line-of-sight (LOS) velocity distribution directly. We assume that the LOS velocity distribution follows a Gaussian function, and the linewidth is defined as two times the full width at half maximum.

\begin{figure}
\centering
\includegraphics[width=11cm]{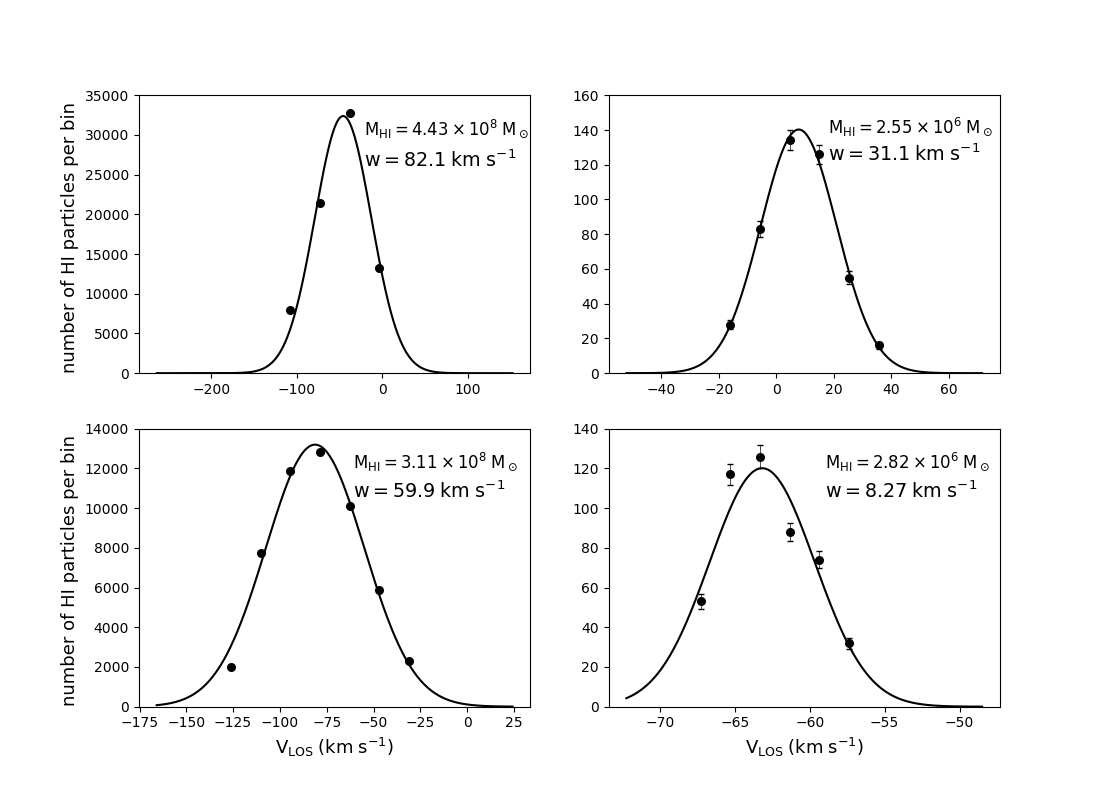}
\caption{LOS velocity distributions for four galaxies. The neutral hydrogen mass and linewidth are also labelled at the top right corner in each panel.}
\label{fig:LW}
\end{figure}

We calculate the linewidth for galaxies with  $M_{\rm HI}>10^5M_\odot$.  In Figure~\ref{fig:LW}, we show the LOS velocity distributions for
four galaxies. The neutral hydrogen mass and linewidth for each galaxy are
also labeled at the top right corner in each panel.   It is seen that the LOS velocity distribution can be fitted well by the Gaussian distribution, and the linewidth increases
with increasing neutral hydrogen mass.

\section{the necessary integration time}
\begin{figure}
\centering
\includegraphics[width=11cm]{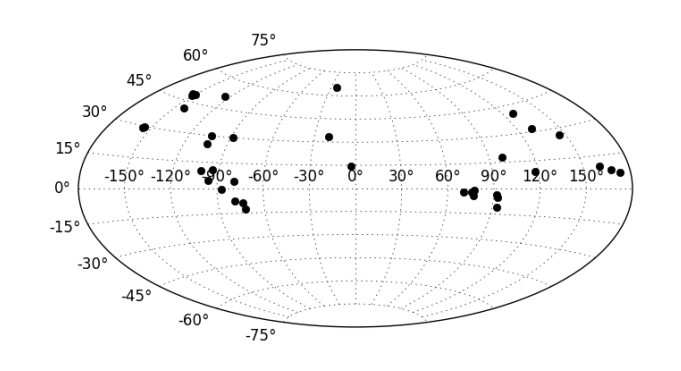}
\caption{Galaxy coordinate distributions in the Local Group from our simulation.  The declination of the galaxies is constrained in the range $[-14^{\circ}, 66^{\circ}]$ by considering the location of FAST.}
\label{fig:coordinate}
\end{figure}

In order to generate mock survey observations, we put M31 at the point with right ascension $0^{\circ} 42' 44''$   and declination $+ 41^{\circ} 16' 19''$. The distance from the Sun to the Galactic center is 8.34 kpc \citep{2014ApJ...783..130R}.
In Figure~\ref{fig:coordinate}, we show the coordinate distribution of galaxies in the local group. The size of the filled circles denotes the relative HI mass for each galaxy.  Since the location of  FAST is  $ 25.647222^{\circ}$N, only galaxies in the declination range $[-14^{\circ}, 66^{\circ}]$ can be detected by FAST.  In our sample, 36 galaxies with $M_{\rm HI}>10^5M_{\odot}$ are located in the FAST field.

\begin{figure}
\centering
\includegraphics[width=11cm]{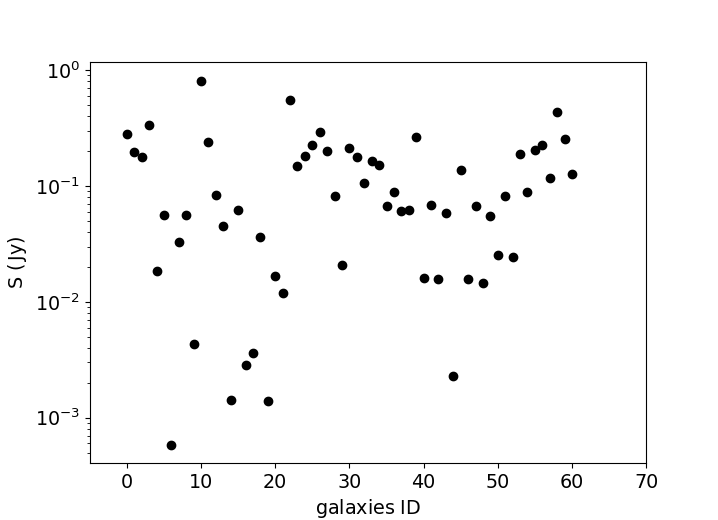}
\caption{Flux of HI from galaxies in the Local Group. The linewidth of galaxies is calculated by using the Gaussian function for each galaxy.}
\label{fig:flux2}
\end{figure}

Combining Equations (1), and (2) with Eq. (3), we know that the detected flux limit depends on the signal-to-noise ratio $S/N$, the frequency bandwidth $\Delta\nu$, the neutral hydrogen mass of the galaxy and the integration time. We also know that the beam size of FAST is three arcminutes and the size of some galaxies in the Local Group is larger than that, therefore, we only consider flux within the FAST beam size. We define that a galaxy can be detected with S/N=6, and adopt $\Delta\nu=1 \rm{MHz}$. 
In Figure ~\ref{fig:flux2}, we show the flux of HI for our sample. 
The detected galaxy number with different integration time is also shown in Table~\ref{tab:int_time}.  It is seen that all satellites with $M_{\rm HI}>10^5\ M_{\odot}$ can be detected by FAST if the integration time is longer than 40 minutes. Limited by the resolution of the \textsc{apostle} simulation, we do not consider galaxies with $M_{\rm HI}<10^5 M_{\odot}$ in this work, but we give an estimation of the integration time for a galaxy with $M_{\rm HI}=10^4\ M_{\odot}$ by using FAST. If this galaxy is 3Mpc from the Earth, then FAST can detect it by observing for 50 minutes.  

In the Local Group, a galaxy's redshift from its peculiar velocity can be comparable with that from the Hubble expansion. In order to calculate the peculiar velocity of each galaxy, we adopt the solar motion $(U_{\odot}, V_{\odot}, W_{\odot})= (9.58, 10.52, 7.01) \rm {km\ s^{-1}}$ \citep{2015ApJ...809..145T}. The observed frequency and flux are affected by the peculiar velocity. The frequency shift can be calculated by

\begin{equation}
\Delta\nu=\nu-\nu_0=(1-\beta)\gamma\nu_0
\end{equation}
where $\beta=v_r/c$, $\gamma=(1-\beta)^2$, $c$ is the velocity of light, and $v_r$ is the radial velocity of a source that emits photons with frequency $\nu_0$ in the rest frame.

\begin{figure}
\centering
\includegraphics[width=11cm]{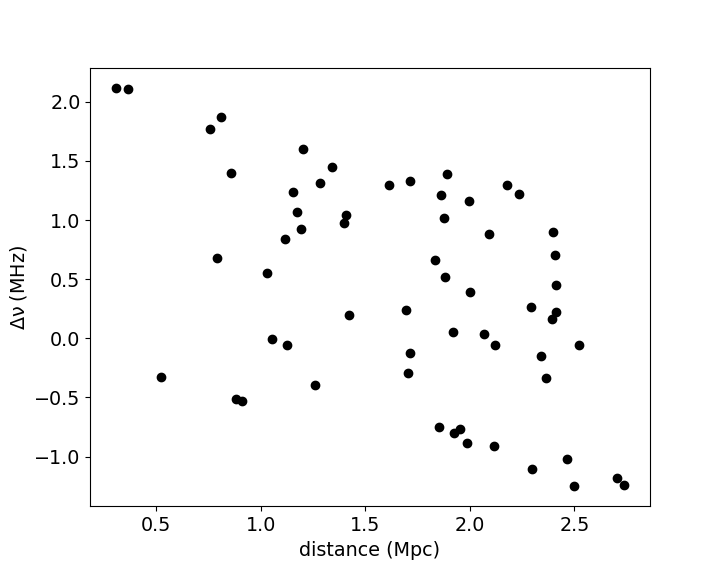}
\caption{Correlation of the frequency drift  with galaxy distance.}
\label{fig:fre_shift}
\end{figure}

 In Figure~\ref{fig:fre_shift}, we show the relation between frequency drift and a galaxy's distance. It is seen that the frequency drift  due to peculiar velocity can be  2MHz for some galaxies.
 For FAST, the $L$ band receiver range is 1.05-1.45 GHz. Therefore, these galaxies still can be detected by FAST. The other effect from peculiar velocity is the observed flux of a galaxy.
 The flux can be increased by $1+z_{pec}$ times, where $z_{pec} $ is the redshift from the peculiar velocity. If $z_{pec}$ is negative, then the observed flux will be smaller than the true flux of the galaxies.
 In our integration time as shown in Table ~\ref{tab:int_time}, this aspect does not affect our results.

\begin{table}
\caption{The integration time for  Local Group galaxies by using FAST.}\label{tab:int_time}
\begin{center}
\begin{tabular}{ccc}
  \hline
 integration time &${\rm number}^a$ &$\rm{number}^b$ \\
  \hline
  t=12\ s &50& 30\\
  t=24\ s&52 & 31  \\
  t=60\ s &54 & 32  \\
  t=300\ s &56 & 32    \\
  t=1200\ s&59 & 34\\
  t=2400\ s & 61 & 36\\
   \hline

\end{tabular}
\end{center}
{\footnotesize
 \noindent
 $^{a}$ the total number of galaxies which can be detected.\\
 $^{b}$ the number of galaxies which can be detected in the FAST field. \\}
\end{table}

\section{summary and discussion}
We have used the high-resolution \textsc{apostle} simulation to study the HI content and dynamical properties of galaxies at the low mass end in the context of the $\Lambda \rm{CDM}$ paradigm. We have compared the HIMF in our simulation  with that  obtained from the ALFALFA survey. We find that our HIMF is consistent with
that of the ALFALFA survey for galaxies with HI mass in the range $10^{6.2}-10^9M_{\odot}$. However, at the low mass end, our HIMF is lower than what is predicted by the ALFALFA survey. The stellar mass function in the simulation is consistent with that of observations if the stellar mass of the galaxies is larger than $10^{6.5}M_{\odot}$. At the low stellar mass end, our simulation predicts more galaxies than what is observed.

We also find that the velocity distribution in a galaxy can be fitted well by a Gaussian function and the linewidth of each galaxy can be obtained directly by this Gaussian distribution.
We study the possibility of detecting these galaxies in the Local Group by using FAST.  We also discuss the effect from peculiar velocity. It is noted that the effect from peculiar velocity is tiny, and can only shift the observed frequency by 1-2 MHz. We find 36 that galaxies with $M_{\rm HI}>10^5{M_\odot}$ in the Local Group can be detected by FAST if the integration time is more than 40 minutes.

In reality, it is difficult to obtain the distance of galaxies in the Local Group by using radio observations, and some high velocity clouds along the LOS can contaminate the observed signal of the HI.  How to remove these fakes is beyond the scope this paper. We will discuss this problem in a future work.

Besides the \textsc{apostle} simulation, similar studies can be done by using semi-analytical models, for example the \cite{2010MNRAS.409..515F}  and \cite{2016MNRAS.458..366L} models can also be used to predict the amount of HI for galaxies in the Local Group.  By using FAST, we can test the validity of theses models, especially at the low HI mass end.

\section*{Acknowledgements}

Thanks Apostle team for sharing the simulation data with us. We acknowledge helpful discussions with Xi Kang, Lincheng Li and Wenkai Hu.We acknowledge the support by the National Science of  Foundation of China (Grant No. 11633004,  11390372, 11303008, 11773034).  JW acknowledges the 973 program grant 2015CB857005 and NSFC grant No. 11373029.



\bibliography{ms}
\bibliographystyle{raa}
\end{document}